\documentclass[aps,a4paper,showpacs,preprintnumbers,amsmath,amssymb,
 floatfix]{revtex4}
\newcommand{\be}{\begin{equation}}
\newcommand{\ee}{\end{equation}}
\newcommand{\bea}{\begin{eqnarray}}
\newcommand{\eea}{\end{eqnarray}}
\newcommand{\ba}{\begin{array}}
\newcommand{\ea}{\end{array}}
\newcommand{\lsim}
{{\;\raise0.3ex\hbox{$<$\kern-0.75em\raise-1.1ex\hbox{$\sim$}}\;}}
\newcommand{\gsim}
{{\;\raise0.3ex\hbox{$>$\kern-0.75em\raise-1.1ex\hbox{$\sim$}}\;}}

\begin{document}
\preprint{
\noindent
\begin{minipage}[t]{2in}
\begin{flushleft}
\end{flushleft}
\end{minipage}
\hfill
\begin{minipage}[t]{2in}
\begin{flushright}
%IISc-CHEP-?/04\\
%\tt{hep-ph/yymmnnn}\\
\vspace*{.2in}
\end{flushright}
\end{minipage}
}
%\draft
%\bibliographystyle{plain}
%\thispagestyle{empty}
%\vspace{-1cm}
%\begin{center}
%\begin{flushright}
%hep-ph/yymmxxx
%\end{flushright}
\title{\bf Critical Behavior of a Relativistic Bose Gas}
\bigskip

\author{P. N. Pandita}
%\email[ ]{pandita@iucaa.ernet.in}
\affiliation{P - 30, North Eastern Hill University,
Shillong 793 022, India}
\begin{abstract}
We show that the thermodynamic behavior of relativistic ideal Bose gas,
recently studied numerically by Grether et al., can be obtained analytically.
Using the analytical results, we obtain the critical behavior of the 
relativistic Bose gas exactly for all the regimes. We show that these
analytical results reduce to those of Grether et al. in different 
regimes of the Bose gas.
\end{abstract}

\vspace{3mm}

%\keywords{Bose-Einstein Condensation, Crtical Bahavior}

\maketitle
%\section{Introduction}

Bose-Einstein condensation~(BEC) of an ideal Bose gas has been a subject of
extensive studies. In particular Bose-Einstein condensation in a relativistic
ideal Bose gas with nonzero chemical potential has been studied by
several authors~\cite{Beckmann:1979jj,AragaoDeCarvalho:1980it,Haber:1981fg,Singh:1982gt, Huang:sm1987}.
In earlier papers~\cite{Beckmann:1979jj,Landsberg:1965pra}  on relativistic ideal Bose condensation, antiboson production 
was not taken into account. The necessity of the antiboson contribution to the thermodyanmics of Bose gas
at relativistic energies was pointed out  in~\cite{Haber:1981fg}, and a high-temperature  expansion
for various thermodynamic quantities  was established and studied. At sufficiently high temperatures, antibosons
are expected to be pair produced in sufficient numbers so that their contribution cannot be neglected.

\medskip

In a recent paper,  Grether et al. studied Bose-Einstein condensation in 
a relativistic ideal 
Bose gas, and calculated its thermodyanamic properties  
numerically~\cite{Grether:2007ur}.
In this note we point out that  the model can be solved exactly 
in a closed form,
including antibosons,  without making any approximations,
and show that an analytic expression for thermodynamic quantities 
can be obtained
at all temperatures and the critical behavior studied  in general. 
The analytical study is physically more transparent,
and sheds light on the connection between relativistic and the non-relativistic
ideal Bose gas by showing that the critical behavior is same for the two 
systems. It is only the overall amplitude which is different for the two 
systems. We also show that our exact results reduce to the 
results of ref. \cite{Grether:2007ur} 
in different regions studied by these authors.

\medskip

The ideal Bose gas is characterized by three basic length scales, namely
the thermal wavelength $\lambda_T,$ the mean interparticle spacing $\overline \lambda,$
and the Compton wavelength $\lambda_C.$ From these, one can obtain two independent ratios
which we consider to be $R_1 = \overline \lambda/\lambda_T,$ and $R_2 = \lambda_C/\lambda_T.$
For quantum~(classical) gas we have $R_1 \ll 1~(\gg 1),$ whereas for $R_2 \ll 1$ and $R_2 \gg 1$
we will have nonrelativistic or ultrarelativistic gas, respectively.  Clearly, it is important to understand 
the connection between different regions, characterized by different length scales, and have  a unified 
treatment for all the regions involved. In this note  we solve the model exactly
in general, which includes all the cases corresponding to all the length scales
and without making a high temperature expansion. We show analytically that the critical
behavior of the model is the same as that of standard
nonrelativistic Bose gas. The difference between nonrelativistic and the 
ultrarelativistic
gas, where antiparticles are important, arises only in the values of critical amplitudes.  

\medskip

The net ``charge'' or number density, assumed positive without loss of generality,  for a relativistic Bose gas
of $N$ bosons and $\overline N$ antibosons, each of mass $m,$ can be written as~(the notation is standard, and
we choose units  $\hbar  = c = k_B = 1, \beta = T^{-1}$; $g$ is the spin degeneracy factor)
\bea
n & = & g \int \frac{d^3 k}{(2\pi)^3} 
      \left[ \frac {1} { {\rm exp} [\beta (k^2 + m^2)^{1/2} 
        - \beta \mu] - 1        }  
        - \frac {1} { {\rm exp} [\beta (k^2 + m^2)^{1/2}
        + \beta \mu] - 1        }      \right],
\label{numberdensity1}
\eea
where we have enclosed the system in a cubical box of
volume $L^3$, with $L$ being the length of an edge in each of the $3$ spatial dimensions.
This can be further written as
\bea
n & = & \frac{m^3 g}{2\pi^2}\int_0^{\infty} dx~~ x^2 ~~ \frac{\sinh (\alpha - \phi)}
{\cosh a(x) - \cosh (\alpha - \phi)} 
 \equiv  \frac{m^3 g}{2 \pi^2} {\cal W}_3(\alpha, \phi),
\label{numberdensity2}
\eea
where $\alpha = \beta m$, $a(x) = \alpha (1 + x^2)^{1/2}$, and
$\phi = -\beta (\mu - m)$. The parameter $\phi$ is defined so as to reduce 
to its nonrelativistic counterpart in that limit. For $d$ spatial dimensions this
generalizes to 
\bea
n & = & a_d^{-1} {\cal W}_d(\alpha, \phi),
\label{numberdensity3}
\eea
where
\bea
a_d & = & 2^{d-1} \pi^{d/2} \Gamma(d/2) m^{-d} g^{-1}, \label {supp1}\\
{\cal W}_d(\alpha, \phi) & = &  \int_0^{\infty} dx~~ x^{d-1}  ~~ \frac{\sinh (\alpha - \phi)} {\cosh a(x) - \cosh (\alpha - \phi)}. 
\label{supp2}
\eea
The results (\ref{numberdensity3}), (\ref{supp1}) and (\ref{supp2})
agree with the result (8) of ref.\cite{Grether:2007ur}.  
Similarly, we can write the energy density as 
\bea
u & = & g \int \frac{d^3 k}{(2\pi)^3}
      \left[ \frac {(k^2 + m^2)^{1/2}} { {\rm exp} [\beta (k^2 + m^2)^{1/2}
              - \beta \mu] - 1        }
	              + \frac {(k^2 + m^2)^{1/2}} 
		         { {\rm exp} [\beta (k^2 + m^2)^{1/2}
		              + \beta \mu] - 1        }      \right],
			      \label{energydensity1}
			      \eea
which can be rewritten as
\bea
u & = & m a_d^{-1} {\cal Y}_d(\alpha,\phi), \label{energydensity2}
\eea
where 
\bea
{\cal Y}_d(\alpha,\phi) & = &  \int_0^{\infty} dx~~ x^{d-1} ~~ \frac{\alpha^{-1} a(x) \cosh (\alpha - \phi) - \sinh (\alpha - \phi)
                         -\alpha^{-1} a(x) e^{-a(x)}}{\cosh a(x) - \cosh (\alpha - \phi)}. \label{supp3}
\eea 
In order to study the  critical behaviour of relativistic Bose gas, it is useful to perform the integrals in (\ref{supp2}) and 
(\ref{supp3}) in a closed form by expanding the integrands in infinite series and introducing modified Bessel functions in the 
resulting integrals. Performing this, we get
\bea
{\cal W}_d(\alpha, \phi) & = & b_d\sum_{r = 1}^\infty r^{-(d'-1)} \sinh (r\alpha - r \phi) K_{d'}(r\alpha),
\label{Bessel1} \\
{\cal Y}_d(\alpha,\phi) & = &  b_d \sum_{r =1}^\infty r ^{-(d'-1)} [\cosh (r\alpha -r \phi) K_{d' + 1}(r\alpha) - \sinh (r\alpha - r\phi) K_{d'}(r\alpha)
                               -(r\alpha)^{-1} \cosh (r\alpha - r \phi) K_{d'}(r\alpha)],
\label{Bessel2}
\eea
where $K_{d'}(r\alpha)$ is a modified Bessel function of order $d'$ and real argument $r\alpha$~\cite{Nieto, AS}, and 
\bea
b_d & = & \pi^{-1/2} 2^{d'}\Gamma(d/2) \overline\alpha^{(d' - 1)}, ~~~~~~~~~~ d' = \frac{1}{2} (d  + 1). \label{supp4}
\eea
We can study the critical behavior of the relativistic Bose gas by studying the variation of $\mu$, or equivalently, $\phi$
as a function of $T.$ By standard argument  of keeping charge density positive, $\mu$ must satisfy the condition $-m \le \mu \le m.$ 
The parameter $\phi$ is, therefore, defined to be positive. From Eqs.(\ref{supp2}) or (\ref{Bessel1}) it is clear
that as $\beta \rightarrow 0, \mu \rightarrow 0,$
or $\phi \rightarrow \beta m$. Decreasing $\beta$, increases $\mu$ until it reaches its limiting value $m$.
The critical temperature is, thus, obtained from  (\ref{numberdensity3}) for $\phi = 0.$  Thus,
\bea
a_d n & = & {\cal W}_d(\alpha_c, 0) = b_{dc} \sum_{r =1}^\infty r^{-(d' - 1)} \sinh (r\alpha_c) K_{d'}(r\alpha_c),
\label{critical1}
\eea
where $b_{dc}$ is the value of $b_d$ at $\alpha_c = \beta_c m = m/T_c.$ Using  the large argument
form of the function $K_\nu(z)$
\bea
K_\nu(z) &  \approx & [\frac{\pi}{2z}]^{1/2} e^{-z} [1  + \frac{4\nu^2 - 1}{8 z} + .......],
\label{largeBessel}
\eea
we see that the   ratio of the successive terms of the series (\ref{critical1}) can be written as
\bea
\frac{u_{r+1}}{u_r} & = & 1 - \frac{d}{2r} + {\cal O}(\frac{1}{r^2}), ~~~~~ r \rightarrow \infty,
\label{ratiotest}
\eea
so that by  ratio test  the series converges for $d > 2.$ 
We, therefore, conclude 
that a nonzero $T_c$ exists. {\em It is important to note that
this result is independent of $\alpha_c (\equiv mc^2/(k_B T_c))$.}

%______

Using  (\ref{numberdensity3}) and (\ref{supp2}) we can determine
the behavior of $\phi$ in the critical region. To do so, we expand  ${\cal W}_d(\alpha, \phi)$ near $\phi = 0$, which 
can be obtained by calculating 
$\partial {\cal W}_d(\alpha, \phi)/\partial\phi$ at $\phi = 0$. We  obtain
\bea
\partial {\cal W}_d(\alpha, \phi)/\partial\phi|_{\phi = 0} & = & -b_d \sum_{r = 1}^\infty r^{-(d' - 2)} \cosh (r\alpha) K_d'(r\alpha).
\label{partialW1}
\eea
This sum converges only for $d > 4.$ The behavior for $2 < d < 4$  as $\phi \rightarrow 0$ can be obtained 
by calculating the derivative
\bea
\partial {\cal W}_d(\alpha, \phi)/\partial\phi & = & -b_d \sum_{r = 1}^\infty r^{-(d' - 2)} \cosh (r\alpha - r \phi)) K_d'(r\alpha),
\label{partialW2}
\eea
from which we get, asymptotically,
\bea
\partial {\cal W}_d(\alpha, \phi)/\partial\phi & \approx & -\frac{1}{2} (2/\alpha)^{d/2} \Gamma(d/2) F_{(d-2)/2}(\phi), \label{asymptot1}
\\
F_n(\phi) & = &  \sum_{r =1}^\infty r^{-n} e^{-r\phi}, \label{asympto2}\\
F_n(\phi) & \approx &  \Gamma(1-n)\phi^{n-1}, ~~~~ n < 1, \phi \rightarrow 0. \label{asympto3}\\
\eea
Putting these results together, we can write
\bea
\partial {\cal W}_d(\alpha, \phi)/\partial\phi| & \approx & -\frac{1}{2} (2/\alpha)^{d/2} \Gamma(d/2) \Gamma(2 - d/2) \phi^{(d-4)/2}, \label{dimension}
\eea
for $2 < d < 4.$ However, the quantity $\partial {\cal W}_d(\alpha, 0)/\partial\alpha$ is calculated to be finite and negative at $\alpha_c.$
We can , therefore, write
\bea
{\cal W}_d(\alpha, 0) & \approx &  {\cal W}(\alpha_c, 0)
                                 - (\alpha - \alpha_c){\cal W}', 
				 \label{expandW1} \\
{\cal W}' & \equiv & -[d{\cal W}_d(\alpha, 0)/d \alpha]_{\alpha = \alpha_c}.\label{expandW2}
\eea
Using (\ref{dimension}) and (\ref{expandW1}) in (\ref{numberdensity3}), we can write
\bea
\phi & \approx & (C^+)^{-1} t^{2/(d-2)}, ~~~~~ 2 < d < 4, ~~~~ t > 0,  \label{final1} \\
C^+ & = & [\frac{2^{d/2} \Gamma(d/2) \Gamma(2-d/2)}{(d-2) \alpha_c^{(d+2)/2} {\cal W}'}]^{2/(d-2)}, \label{final2}
\eea
where we have written
\bea
t & =& (T - T_c)/T_c. \label{final3}
\eea
From the result (\ref{final1})  we note that the critical exponent is
$2/(2-d),$ which is the same as that of the nonrelativistic Bose gas. Thus, the critical behavior of the gas is same
in all regions, whether relativistic or nonrelativistic~\cite{GB}. The only difference
relates to the  amplitude, the overall factor multiplying
$[(T - T_c)/T_c]^{2/(d-2)}$ in (\ref{final1}). 
Finally, turning to the behavior
for  $T < T_c$, we note that $\phi =0 $ for  $T < T_c$.

\bigskip

In order to see the connection of  our general analytical results 
with the  results of ref. \cite{Grether:2007ur}, we consider the 
nonrelativistic~(NR) and the 
ultrarelativistic~(UR) limit for the number density and the energy density
of the Bose gas. These correspond to $\alpha_c \gg 1 $ and  $\alpha_c \ll 1, $
respectively. From Eq.(\ref{critical1}), we have
\bea
{\cal W}_d(\alpha_c, 0) & = & b_{dc} \sum_{r =1}^\infty r^{-(d' - 1)} 
                               \sinh (r\alpha_c) K_{d'}(r\alpha_c).
\label{limitcritical1}
\eea
Using (\ref{largeBessel}), we get for the nonrelativistic case 
\bea
{\cal W}_d(\alpha_c, 0) & = & \frac{1}{2}\left[\frac{2}{\alpha_c}\right]^{d/2}
                              \Gamma(d/2) \zeta(d/2), ~~~~\alpha_c \gg 1.
			      \label{NR1}
\eea
Using the fact that ${\cal W}_d(\alpha_c, 0) = a_d n,$ and solving (\ref{NR1})
for the critical temperature $\alpha_c =  m/T_c,$ we get
\bea
T_c & = & \frac{2\pi}{m} \left[\frac{n}{g \zeta(d/2)}\right]^{2/d}, \label{NR2}
\eea
in the nonrelativistic region, 
which is the result (5) of ref.\cite{Grether:2007ur}. Using 
(\ref{energydensity2}), (\ref{supp3}), and (\ref{Bessel2}), we can write
for the energy density in the nonrelativistic region
\bea
u & = & m a_d^{-1} {\cal Y}_d(\alpha_c, 0), \label{NR3}
\eea
where
\bea
{\cal Y}_d(\alpha_c,0) & = & \frac{1}{2\alpha_c}
                            \left[ \frac{2}{\alpha_c} \right]^{d/2}
			    \Gamma(\frac{d+2}{2})\zeta(d/2), ~~~~~ 
			    \alpha_c\gg 1.
			    \label{NR4}
\eea
On the other hand for the ultrarelativistic limit $\alpha_c \ll 1,$ we can 
use the expansion~\cite{AS}
\bea
K_{\nu}(z) &\approx & \frac {2^{\nu -1} \Gamma(\nu)}{z^{\nu}}, \label{UR1}
\eea
for $z \ll 1$ to obtain
\bea
{\cal W}_d(\alpha_c, 0) & = & \frac{2}{\alpha_c^{d-1}}\Gamma(d)\zeta(d-1),
                            ~~~~~~~~ \alpha_c \ll1   \label{UR2} 
\eea
which, together with the result (\ref{critical1}), can be solved for the 
critical temperature to obtain
\bea
T_c & = & \left [\frac{n (2\pi)^d \Gamma(d/2)}
                       {4 g m\pi^{d/2} \Gamma(d) \zeta (d-1)}
		       \right]^{1/(d-1)}, ~~~~~~~~\alpha_c \ll 1,
		       \label{UR3}
\eea
which is the same as the result (9) of ref.~\cite{Grether:2007ur}.
Finally, in the UR limit the energy density can be written as
\bea
u & = & m a_d^{-1} {\cal Y}_d(\alpha_c, 0), \label{UR4}\\
{\cal Y}_d(\alpha_c,0) & = & \frac{2}{\alpha_c} (d-1) \Gamma(d) \zeta(d-1),
			    ~~~~~ \alpha_c\ll 1. \label{UR5} 
\eea
In summary we have solved the ideal, relativistic Bose gas, in  a closed form,
and have shown that the critical behavior of the gas is same as for 
the standard nonrelativistic Bose gas. Our general results are applicable to 
all the regions of the Bose gas. We have shown that our general results
reduce to the known results in the nonrelativistic and the 
ultrarelativistic regions. We have, thus, obtained a unified treatment
of the Bose gas which is applicable in all regions of the parameter space.

\bigskip

\bigskip

The author thanks the Inter University Centre for Astronomy and 
Astrophysics, Pune, India for hospitality while this work was completed.
This work is supported by the J. C. National Bose Fellowship of the
Department of Science and Technology, and by the
Council of Scientific and Industrial Research, India under
Project No. (03)(1220)/12/EMR-II.

\bigskip

\bigskip

%%%%%%%%%%%%%%%%%%%%%%%%%%%%%%%%%%%%%%%%%%%%%%%%%%%%%%

\end{document}